\documentclass[article,preprint,groupedaddress]{revtex4}
\usepackage{epsfig}
\usepackage{graphicx}

\begin{document}

\title{Self-gravitating field configurations: The role of the energy-momentum trace}
\author{Shahar Hod}
\address{The Ruppin Academic Center, Emeq Hefer 40250, Israel}
\address{}
\address{The Hadassah Institute, Jerusalem 91010, Israel}
\date{\today}

\begin{abstract}

\ \ \ Static spherically-symmetric matter distributions whose
energy-momentum tensor is characterized by a non-negative trace
are studied analytically within the framework of general relativity.
We prove that such field configurations are necessarily highly
relativistic objects. In particular, for matter fields with
$T\geq\alpha\cdot\rho\geq0$ (here $T$ and $\rho$ are respectively
the trace of the energy-momentum tensor and the energy density of
the fields, and $\alpha$ is a non-negative constant), we obtain the
lower bound $\text{max}_r\{2m(r)/r\}>(2+2\alpha)/(3+2\alpha)$ on the
compactness (mass-to-radius ratio) of regular field configurations.
In addition, we prove that these compact objects necessarily possess
(at least) {\it two} photon-spheres, one of which exhibits {\it
stable} trapping of null geodesics. The presence of stable
photon-spheres in the corresponding curved spacetimes indicates that
these compact objects may be nonlinearly unstable. We therefore
conjecture that a negative trace of the energy-momentum tensor is a
{\it necessary} condition for the existence of stable, soliton-like
(regular) field configurations in general relativity.
\end{abstract}
\bigskip
\maketitle

\section{Introduction}

Nonlinear solitons have a long and broad history in science. These
regular particle-like configurations find applications in many areas
of physics, such as: general relativity \cite{Vol}, string theory
\cite{Duf}, condensed matter physics \cite{Bis}, nonlinear optics
\cite{Agg}, and astrophysics \cite{Bos}.

Let us denote by $T^{\mu}_{\nu}$ the energy-momentum tensor of the
matter fields which compose a nonlinear static soliton. A simple
argument \cite{Vol,Des} then reveals that, in flat space, the sum of
the principal pressures, $\Sigma_i p_i$ (here $p_i=T^{i}_{i}$
), cannot have a fixed sign throughout the body volume.
This can be seen from the conservation law
${\partial_jT^{j}_{i}=0}$, which implies that the spatial components
of the energy-momentum tensor satisfy \cite{Vol,Des}
\begin{equation}\label{Eq1}
\int_{R^3}T_{ij}d^3x=0\  .
\end{equation}
The volume integral (\ref{Eq1}) has a simple physical meaning: it
states that the total stresses must balance in a static matter
distribution \cite{Vol,Des}. The relation (\ref{Eq1}) then implies
that the sum of the principal pressures must switch signs somewhere
inside the volume of the extended body. In particular, no regular
static matter distributions exist with $\Sigma_i p_i>0$ throughout
the entire space. Such systems are of a purely repulsive nature and
thus the force balance is impossible \cite{Vol,Des}.

Although for purely repulsive matter fields (with $\Sigma_i p_i>0$
throughout the body volume) in {\it flat} space the force balance is
impossible, the situation may change in {\it curved} spacetimes
(that is, in the presence of gravity). This fact is nicely
demonstrated by the existence of globally regular particle-like
solutions of the coupled Einstein-Yang-Mills field equations
\cite{BarMic}. These non-linear solitons describe extended objects
in which the repulsive nature of the matter field \cite{Notep} is
balanced by the attractive nature of gravity.

\section{The trace of the energy-momentum tensor}

We have seen that any flat-space static matter distribution must be
characterized by the relation $\Sigma_i p_i<0$ in some part of it.
Denoting by $\rho>0$ \cite{Noteweak} the energy-density of the
matter fields, one concludes that the trace of the energy-momentum
tensor,
\begin{equation}\label{Eq2}
T\equiv-\rho+\Sigma_i p_i\  ,
\end{equation}
must also be negative in this part of the system volume. Thus, a
negative trace of the energy-momentum tensor, at least in some part
of the system volume, is a necessary condition for the existence of
static regular matter distributions in flat spacetimes.

However, it should be emphasized that this conclusion no longer
holds true in curved spacetimes. In particular, static soliton-like
field configurations which are characterized by a non-negative trace
(throughout the {\it entire} configuration's volume) do exist. The
gravitating Einstein-Yang-Mills solitons \cite{BarMic}, which are
characterized by the identity $T=0$, are a well-known example for
such regular particle-like configurations.

The main goal of the present paper is to analyze, within the
framework of general relativity, the physical properties of regular
self-gravitating field configurations whose energy-momentum tensor
is characterized by a non-negative trace \cite{Notesub}. The rest of
the paper is organized as follows: In Sec. III we shall describe our
physical system. In particular, we shall formulate the Einstein
field equations in terms of the trace of the energy-momentum tensor.
In Sec. IV we shall prove that matter configurations which are
characterized by a non-negative energy-momentum trace are
necessarily highly relativistic objects. In particular, we shall
derive a lower bound on the compactness (mass-to-radius ratio
\cite{Bond1,Hodm}) of these extended physical objects. In Sec. V we
shall prove that the curved spacetime geometries which describe
these self-gravitating objects necessarily possess (at least) two
photon-spheres, compact hypersurfaces on which massless particles
can follow null circular geodesics. We shall show that one of these
photon-spheres exhibits stable trapping of the null circular
geodesics. We conclude in Sec. VI with a summary of the main
results.

\section{Description of the system}

We study static spherically symmetric matter configurations in
asymptotically flat spacetimes. The line element describing the
spacetime geometry takes the following form in Schwarzschild
coordinates \cite{Hodm,Nun,Chan,CarC}
\begin{equation}\label{Eq3}
ds^2=-e^{-2\delta}\mu dt^2 +\mu^{-1}dr^2+r^2(d\theta^2 +\sin^2\theta
d\phi^2)\  .
\end{equation}
The metric functions $\delta(r)$ and $\mu(r)$ in (\ref{Eq3}) depend
on the Schwarzschild areal coordinate $r$. Regularity of the matter
configurations at the center requires
\begin{equation}\label{Eq4}
\mu(r\to 0)=1+O(r^2)\ \ \ {\text{and}}\ \ \ \delta(0)<\infty\  .
\end{equation}
In addition, asymptotically flat spacetimes are characterized by
\begin{equation}\label{Eq5}
\mu(r\to\infty) \to 1\ \ \ {\text{and}}\ \ \ \delta(r\to\infty) \to
0\ .
\end{equation}

The fields that compose the matter configurations are characterized
by an energy-momentum tensor $T^{\mu}_{\nu}$. The Einstein
equations, $G^{\mu}_{\nu}=8\pi T^{\mu}_{\nu}$, are given by
\cite{Hodm,Nun,Noteunits}
\begin{equation}\label{Eq6}
\mu'=-8\pi r\rho+(1-\mu)/r\  ,
\end{equation}
and
\begin{equation}\label{Eq7}
\delta'=-4\pi r(\rho +p)/\mu\  ,
\end{equation}
where $T^{t}_{t}=-\rho$, $T^{r}_{r}=p$, and
$T^{\theta}_{\theta}=T^{\phi}_{\phi}=p_T$ are respectively the
energy density, the radial pressure, and the tangential pressure of
the fields \cite{Bond1}, and a prime denotes differentiation with
respect to $r$.

The gravitational mass $m(r)$ contained within a sphere of radius
$r$ is given by \cite{Notemas}
\begin{equation}\label{Eq8}
m(r)=\int_{0}^{r} 4\pi x^{2} \rho(x)dx\  .
\end{equation}
For the total mass of the configuration to be finite, the energy
density $\rho$ should approach zero faster than $r^{-3}$ at spatial
infinity:
\begin{equation}\label{Eq9}
r^3\rho(r)\to 0\ \ \ \text{as} \ \ \ r\to\infty\  .
\end{equation}

Substituting the Einstein field equations (\ref{Eq6}) and
(\ref{Eq7}) into the conservation equation
\begin{equation}\label{Eq10}
T^{\mu}_{r ;\mu}=0\  ,
\end{equation}
one finds
\begin{eqnarray}\label{Eq11}
p'(r)= {{1} \over {2\mu r}}\big[{\cal N}(\rho+p)+2\mu T-8\mu p\big]\
\end{eqnarray}
for the pressure gradient, where
\begin{equation}\label{Eq12}
T\equiv-\rho+p+2p_T\
\end{equation}
is the trace of the energy-momentum tensor, and
\begin{equation}\label{Eq13}
{\cal N}(r)\equiv 3\mu-1-8\pi r^2p\  .
\end{equation}
Below we shall analyze the spatial behavior of the pressure function
$P(r) \equiv r^{2}p(r)$, whose gradient is given by [see Eq.
(\ref{Eq11})]
\begin{eqnarray}\label{Eq14}
P'(r)= {{r} \over {2\mu}}\big[{\cal N}(\rho+p)+2\mu T-4\mu p\big]\ .
\end{eqnarray}

We shall assume that the matter fields satisfy the following
conditions:
\newline
(1) The dominant energy condition \cite{HawEl}. This means that the
energy density bounds the pressures:
\begin{equation}\label{Eq15}
\rho\geq |p|,|p_T|\geq 0\  .
\end{equation}
(2) The trace of the energy-momentum tensor is non-negative.
Specifically, we shall assume that the trace is bounded from below
by
\begin{equation}\label{Eq16}
T \geq \alpha\cdot\rho\geq 0\  ,
\end{equation}
where $\alpha\geq0$ is a constant. Note that Eqs. (\ref{Eq12}) and
(\ref{Eq15}) imply $T\leq 2\rho$, an inequality which restricts the
value of $\alpha$ to the regime \cite{Notealph,Zelfm}:
\begin{equation}\label{Eq17}
0\leq \alpha\leq 2\  .
\end{equation}
From Eqs. (\ref{Eq12}), (\ref{Eq15}), and (\ref{Eq16}) one also
finds
\begin{equation}\label{Eq18}
p_T={1\over 2}[T+(\rho-p)]\geq 0\  .
\end{equation}

\section{Lower bound on the compactness of the matter distributions}

In the present section we shall derive a lower bound on the
compactness, $\text{max}_r\{2m(r)/r\}$, of the regular field
configurations. To that end, we shall first analyze the behavior of
the pressure function $P(r)$ in the asymptotic regimes $r\to 0$ and
$r\to\infty$:
\newline
(1) From Eqs. (\ref{Eq4}) and (\ref{Eq11}) one finds
$p'(r)=2(p_T-p)/r$ as $r\to 0$. Regularity of $p(r)$ therefore
requires $p(0)=p_T(0)\geq0$ [see Eq. (\ref{Eq18})], which implies
\cite{Noteinad}
\begin{equation}\label{Eq19}
P(r\to 0)\to 0^+\  .
\end{equation}
\newline
(2) From Eqs. (\ref{Eq5}) and (\ref{Eq14}) one finds $P'(r)\simeq
2rp_T$ as $r\to \infty$, which implies \cite{Notept}
\begin{equation}\label{Eq20}
P'(r\to\infty)\to 0^+\  .
\end{equation}
In addition, from Eqs. (\ref{Eq9}) and (\ref{Eq15}) one learns that
$p(r)$ should approach zero faster than $r^{-3}$ at spatial
infinity, which implies \cite{Notefb}
\begin{equation}\label{Eq21}
P(r\to\infty)\to 0^-\  .
\end{equation}

Inspection of Eqs. (\ref{Eq19}) and (\ref{Eq21}) reveals that the
pressure function $P(r)$ must switch signs at some intermediate
point, $r=r_0$, such that:
\begin{equation}\label{Eq22}
P(r=r_0)=0\ \ \ \text{and}\ \ \ P'(r=r_0)\leq 0\  .
\end{equation}
Note that the negativity of the pressure function $P(r)$ in some
part of the spacetime [see Eq. (\ref{Eq21})] implies that $0\leq T<
\rho$ in this spacetime region, an inequality which further
restricts the value of $\alpha$ to the regime
\begin{equation}\label{Eq23}
0\leq \alpha<1\  .
\end{equation}

Taking cognizance of Eqs. (\ref{Eq13}), (\ref{Eq14}), (\ref{Eq15}),
(\ref{Eq16}), and (\ref{Eq22}), one concludes that
$\mu(3+2\alpha)-1\leq 0$ at $r=r_0$, which yields the upper bound
\begin{equation}\label{Eq24}
\mu\leq {{1}\over{3+2\alpha}}\
\end{equation}
at $r=r_0$, or equivalently \cite{Notemas}
\begin{equation}\label{Eq25}
{\text{max}}_r\Big\{{{2m(r)} \over r}\Big\}\geq{{2+2\alpha}\over
{3+2\alpha}}\ .
\end{equation}
Note that the larger is the value of $\alpha$, the stronger is the
lower bound (\ref{Eq25}) \cite{Noteeym}.

\section{Photonspheres of the regular compact objects}

One of the most remarkable features of black-hole spacetimes is the
existence of photonspheres, closed hypersurfaces on which massless
particles (null rays) can orbit the central black hole. In the
present section we shall prove that compact {\it regular} objects
whose energy-momentum tensor is characterized by a non-negative
trace [see Eq. (\ref{Eq16})] necessarily possess (at least) two
photonspheres.

To that end, we shall first follow the analysis of
\cite{Chan,CarC,Hodhair} in order to determine the radius(es) of the
null circular geodesic(s) in the regular spacetime (\ref{Eq3}). The
geodesic motions of test particles in the curved spacetime
(\ref{Eq3}) are governed by the characteristic equation
\cite{Chan,CarC,Hodhair}
\begin{equation}\label{Eq26}
E^2-V_r\equiv \dot
r^2=\mu\Big({{E^2}\over{e^{-2\delta}\mu}}-{{L^2}\over{r^2}}-\epsilon\Big)\
,
\end{equation}
where a dot denotes differentiation with respect to proper time.
Here $V_r$ is the effective radial potential governing the motion of
the test particles with $\epsilon=0$ for null geodesics and
$\epsilon=1$ for timelike geodesics, and $\{E,L\}$ are constants of
the motion reflecting the independence of the metric (\ref{Eq3}) on
both $t$ and $\phi$.

Circular geodesics are characterized by $E^2=V_r$ and $V^{'}_r=0$
[$\dot r^2=(\dot r^2)'=0$] \cite{Chan,CarC,Hodhair}. These two
equations yield the relation
\begin{equation}\label{Eq27}
2e^{-2\delta}\mu-r(e^{-2\delta}\mu)^{'}=0\
\end{equation}
for the null circular geodesics of the spacetime geometry
(\ref{Eq3}). Substituting the Einstein field equations (\ref{Eq6})
and (\ref{Eq7}) into (\ref{Eq27}), one obtains the characteristic
equation
\begin{equation}\label{Eq28}
{\cal N}(r=r_{\gamma})=0
\end{equation}
for null circular geodesics in the curved spacetime. Taking
cognizance of Eqs. (\ref{Eq4}), (\ref{Eq5}), (\ref{Eq9}),
(\ref{Eq13}), and (\ref{Eq15}), one finds
\begin{equation}\label{Eq29}
{\cal N}(r=0)=2\ \ \ \text{and}\ \ \ {\cal N}(r\to\infty)\to 2\
\end{equation}
at the two boundaries of the spacetime.

Inspection of Eqs. (\ref{Eq19}) and (\ref{Eq21}) reveals that there
must be a finite interval, $(r_a,r_b)$, in which the pressure
function $P(r)$ is negative and decreasing:
\begin{equation}\label{Eq30}
P(r_a<r<r_b)<0\ \ \ \text{and}\ \ \ P'(r_a<r<r_b)<0\  .
\end{equation}
Taking cognizance of Eqs. (\ref{Eq14}), (\ref{Eq15}), (\ref{Eq16}),
and (\ref{Eq30}), one deduces that
\begin{equation}\label{Eq31}
{\cal N}(r_a<r<r_b)<0
\end{equation}
in this interval. From Eqs. (\ref{Eq29}) and (\ref{Eq31}) one
concludes that there must be (at least) two radii, $r_{\gamma 1}$
and $r_{\gamma 2}$, which are characterized by
\begin{equation}\label{Eq32}
{\cal N}(r_{\gamma 1})={\cal N}(r_{\gamma 2})=0\  .
\end{equation}
These two radii correspond to {\it two} photon-spheres which
characterize our regular curved spacetimes.

It is worth emphasizing that the regular spacetime (\ref{Eq3}) may
be characterized by more than two photon-spheres. We shall
henceforth denote by $r_{\gamma 1}$ and $r_{\gamma 2}$ the innermost
and the outermost radii of these photonspheres, respectively. Note
that this implies [see Eqs. (\ref{Eq29}) and (\ref{Eq31})]
\begin{equation}\label{Eq33}
{\cal N}'(r=r_{\gamma 1})\leq0\ \ \ \text{and}\ \ \ {\cal
N}'(r=r_{\gamma 2})\geq0\  ,
\end{equation}
where the derivative ${\cal N}'(r=r_{\gamma})$ can be expressed in
the form [see Eqs. (\ref{Eq6}), (\ref{Eq11}), and (\ref{Eq13})]
\begin{equation}\label{Eq34}
{\cal N}'(r=r_{\gamma})={{2}\over {r_{\gamma}}}\big[1-8\pi
r^2_{\gamma}(\rho+p_T)\big]\  .
\end{equation}

The stability/instability properties of the characteristic circular
geodesics are determined by the second spatial derivative of the
effective radial potential \cite{Chan,CarC}: unstable circular
geodesics are characterized by $V^{''}_r<0$, whereas stable circular
geodesics are characterized by $V^{''}_r>0$. From Eqs. (\ref{Eq6}),
(\ref{Eq7}), and (\ref{Eq26}) one obtains
\begin{equation}\label{Eq35}
V^{''}_r(r=r_{\gamma})=-{{2E^2e^{2\delta}}\over{\mu
r^2_{\gamma}}}\big[1-8\pi r^2_{\gamma}(\rho+p_T)\big]\
\end{equation}
for null circular geodesics. Taking cognizance of Eqs. (\ref{Eq33}),
(\ref{Eq34}), and (\ref{Eq35}), one finds
\begin{equation}\label{Eq36}
V^{''}_r(r=r_{\gamma 1})\geq0\ \ \ \text{and}\ \ \
V^{''}_r(r=r_{\gamma 2})\leq0\  .
\end{equation}
One therefore concludes that the inner photonsphere is stable,
whereas the outer photonsphere is unstable.

\section{Summary and physical implications}

In this paper we have analyzed the physical properties of
self-gravitating field configurations whose energy-momentum tensor
is characterized by a non-negative trace: $T\geq
\alpha\cdot\rho\geq0$. The main results obtained in this paper and
their physical implications are:
\newline
(1) It has been shown that, regularity of the field configurations
sets an upper bound on the magnitude of the energy-momentum trace
[see Eq. (\ref{Eq23})]:
\begin{equation}\label{Eq37}
T<\rho\  .
\end{equation}
This bound excludes, for example, the existence of high density
fluid matter configurations in which the inter-particle interactions
are mediated by massive vector fields (such fluids are characterized
by $T\to 2\rho$ \cite{Zelfm}).
\newline
(2) It has been shown that these matter configurations are
necessarily highly relativistic objects. In particular, we obtained
the lower bound
\begin{equation}\label{Eq38}
{\text{max}}_r\Big\{{{2m(r)} \over r}\Big\}\geq{{2+2\alpha}\over
{3+2\alpha}}\
\end{equation}
on the mass-to-radius ratio (compactness) of these regular field
configurations.
\newline
(3) We have proved that regular objects whose energy-momentum tensor
is characterized by a non-negative trace necessarily possess (at
least) two photonspheres, one of which exhibits {\it stable}
trapping of null circular geodesics [see Eq. (\ref{Eq36})].

In this respect, it is interesting to mention an important recent
result by Keir \cite{Keir}: It was shown \cite{Keir} that in
spherically symmetric, asymptotically flat regular spacetimes which
exhibit stable trapping of null geodesics, linear perturbation
fields cannot decay faster than logarithmically. As emphasized in
\cite{Keir,Noteins}, this result indicates that the corresponding
compact objects may be nonlinearly unstable.

Our results therefore suggest that, while gravity may allow for the
existence of regular field configurations whose energy-momentum
tensor is characterized by a non-negative trace \cite{Noterec},
these highly compact objects are probably nonlinearly unstable
\cite{Notetwo}. It is therefore tempting to conjecture that, within
the framework of classical general relativity, a negative trace of
the energy-momentum tensor (at least in some part of the system
volume) is a necessary condition for the existence of stable,
particle-like (regular) field configurations.

\bigskip
\noindent {\bf ACKNOWLEDGMENTS}

This research is supported by the Carmel Science Foundation. I thank
Yael Oren, Arbel M. Ongo and Ayelet B. Lata for stimulating
discussions.


\begin{thebibliography}{99}
\bibitem{Notep} Note that the Yang-Mills field is characterized by $\Sigma_i p_i>0$.
\bibitem{Noteweak} We assume that the matter fields respect the weak
energy condition.
\bibitem{Noteunits} We shall use natural units in which $G=c=1$.
\bibitem{Notealph} Note that the $\alpha=0$ case corresponds, for example, to the
Einstein-Yang-Mills solitons \cite{BarMic}. The $\alpha\to 2\ (T\to
2\rho)$ case corresponds, for example, to high density fluid matter
in which the inter-particle interactions are mediated by massive
vector fields, see \cite{Zelfm} for details.
\bibitem{Noteinad} In addition, from Eqs. (\ref{Eq4}) and (\ref{Eq14}) one finds $P'(r)\simeq 2rp_T$ as
$r\to 0$, which implies [see Eq. (\ref{Eq18})] $P'(r\to 0)\to 0^+$.
\bibitem{Notept} Here we have used Eq. (\ref{Eq18}) together with the fact that
$p_T(r)$ should approach zero faster than $r^{-3}$ at spatial
infinity [see Eqs. (\ref{Eq9}) and (\ref{Eq15})].
\bibitem{Notefb} Here we have also used Eq. (\ref{Eq20}) in order to deduce the sign of $P(r\to\infty)$.
\bibitem{Noteeym} Note that the $\alpha=0\ (T=0)$ case was studied in \cite{Hodm},
where the lower bound ${\text{max}}_r\{{2m(r)}/r\}\geq 2/3$ was
obtained. It should be noted that in \cite{Hodm} we assumed that the
energy density $\rho$ of the matter distribution goes to zero
asymptotically faster than $r^{-4}$. We would like to emphasize that
in the present paper we provide a more {\it general} proof of the
lower bound (\ref{Eq25}) which avoids using this restrictive
assumption.
\bibitem{Notetwo} It is worth emphasizing again that, we have proved that such regular
objects (with $T\geq 0$) are necessarily characterized by the
presence of {\it stable} null circular geodesics in the
corresponding curved spacetimes, see Eq. (\ref{Eq36}).
\end{thebibliography}
\end{document}